\documentclass{aa}  

\usepackage{graphicx}
\usepackage{amsmath}
\usepackage{txfonts}
\usepackage{longtable}
\usepackage{natbib}
\setcitestyle{round}

\DeclareMathSizes{10}{12}{9}{7}

\usepackage{lipsum}
\usepackage{lscape}             
\usepackage{placeins}
\usepackage{capt-of}

\makeatletter
\@fleqnfalse
\makeatother

\usepackage{etoolbox}

\AtBeginEnvironment{equation}{\Large}
\AtBeginEnvironment{align}{\Large}
\AtBeginEnvironment{gather}{\Large}

\makeatletter
\renewcommand{\linenumbers}{}
\renewcommand{\nolinenumbers}{}
\makeatother

\begin{document}

   \title{A Physical Classification of Exoplanet Thermal Environments: Stellar Irradiation versus Tidal Heating}

   \author{Daniel Fadrique Barbero\inst{1}}
        
   \institute{Independent Researcher, "astrodanielfb@gmail.com"}
   
   \date{}

  \abstract{
In this study, we introduce a physical framework to analyse and classify the thermal regimes governed by tidal heating and stellar irradiation. Although all planetary systems are exposed to stellar radiation, this source is not always the dominant energy mechanism. This study is motivated by the lack of a physical framework that examines tidal heating in cases where this phenomenon dominates over stellar irradiation. We develop a reproducible physical approach that allows us to classify the relative contribution of both fluxes in a population of exoplanets, identifying the most relevant physical mechanisms that determine the thermal regime. We apply this method to a population of approximately 2000 exoplanets. This framework is centred on the dimensionless parameter \(\Lambda=F_{\mathrm{abs}}/F_{\mathrm{tide}}\), which quantifies the relative contribution of each flux. Our results show that most planetary thermal environments are dominated by \(F_{\mathrm{abs}}\), although there is a significant fraction of systems in which the tidal flux dominates. We identify a physical boundary at \(\Lambda=1\) that defines a regime in which both fluxes are comparable. We identify the semi-major axis \(a\) and the eccentricity \(e\) are the parameters that most influence the tidal flux. This framework provides a transparent and physically motivated tool for characterising planetary thermal environments and exploring the physical trends governing exoplanet populations. 
  }


   \maketitle

\section{Introduction}

One of the key factors for potential habitability is the thermal environment of exoplanets \citep{seager2010, Pierrehumbert2011, Kopparapu2014} which is characterised by how they receive energy and how they process it \citep{cowan2011}. In most observed planetary systems, stellar radiation is the primary source of energy; however, when certain conditions are met, tidal heating arises \citep{ogilvie2014}, which may be secondary, comparable or dominant \citep{jackson2008, Barnes2009, Henning2009} to the incident stellar flux depending on the system’s stellar, planetary and orbital parameters \citep{rasio1996, laskar1993}. 

Both heat fluxes can coexist within the same thermal regime, competing to dominate it. In this context, a fundamental question arises: in systems experiencing both energy sources, which of them dominates the thermal environment and how can the two phenomena be compared?  

In the current literature, there are many articles studying stellar irradiation or tidal heating \citep{seager2010, jackson2008, Henning2009, Barnes2009}. However, there is no simple, reproducible and clear physical framework that directly compares them by classifying in which cases each flux dominates. 

In this article, we propose a framework based on the dimensionless parameter \(\Lambda\), which allows us to quantify and classify the thermal regimes dominated by each energy source  in a given thermal environment. 

In our analysis, we use a sample of approximately 2000 exoplanets, which we organise within a simple and reproducible physical framework. This allows us to classify them according to \(\Lambda\) and to study the orders of magnitude of the stellar, planetary and orbital parameters of each planetary system.  

This approach highlights a fundamental hierarchy and provides a physically interpretable classification of the physical mechanisms governing tidal heating and, consequently, the classification of planetary thermal environments.

\section{Theoretical Framework}

This article analyses the main heat fluxes that determine the thermal regime of an exoplanet. Stellar irradiation, which is always present, coexists with tidal dissipation in planetary systems where \(e>0\). The main thermal source for exoplanets is the incident stellar flux \citep{guillot2010}, although under certain conditions it competes with the tidal flux to dominate the planetary thermal environment, revealing a physical boundary where one flux dominates the other or both are comparable. This highlights the need for a physical parameter for systematic classification. 

\subsection{The Thermal Environment of Exoplanets}

The thermal environment of an exoplanet is characterised by the balance between the energy it receives and the energy it emits:  

\begin{equation} 
F_{\mathrm{in}}=F_{\mathrm{out}} 
\end{equation} 

We assume radiative equilibrium as a first stationary approximation. To determine the radiative equilibrium of a planetary system, we analyse its main thermal source, the incident stellar flux \(F_{\mathrm{abs}}\). Although this flux is dominant in many of the observed regimes, the thermal phenomenon produced by tides, \(F_{\mathrm{tide}}\), is relevant for characterising the thermal environment of an exoplanet. 

\subsection{The Incident Stellar Flux}

Stellar radiation, or incident stellar flux, is the dominant factor for most observed exoplanets. For systems with circular orbits where \(e=0\), we can attribute radiative equilibrium exclusively to the flux \(F_{\mathrm{abs}}\) \citep{seager2010}. 

\begin{equation} 
F_{\mathrm{abs}}=\frac{L_{\star}}{4 \pi a^{2}}(1-A) 
\end{equation} 

Where \(A\) is the Bond Albedo, a dimensionless parameter that quantifies the fraction of incident energy reflected back into space. 

\subsection{The Tidal Heating Flux}

Planetary systems that satisfy the condition \(e>0\) and have orbits with a small semi-major axis \(a\) will experience tidal forces \(F_{\mathrm{tide}}\) alongside stellar radiation. The tidal dissipation power in \(\text{W}\) is calculated using the formula \citep{peale1979, Hut1981, jackson2008}: 

\begin{equation} 
\dot{E}_{\mathrm{tide}}=\frac{21}{2}\frac{k_{2}}{Q}\frac{GM_{\star}^{2}R_{p}^{5}}{a^{6}}ne^{2} 
\end{equation} 

Where the fraction \(\frac{21}{2}\) represents the numerical factor of the tidal model, \(k_{2}\) is an internal parameter of the planet that quantifies the planet’s deformation in response to the gravitational potential, \(Q\) is the tidal quality factor \citep{GOLDREICH1966375, Murray_Dermott2000} that quantifies the efficiency of energy dissipation per cycle, the smaller \(Q\) is, the greater the dissipation the planet will experience, \(G\) is the gravitational constant and \(n\), which establishes the timescale of the tidal forces, is calculated as \(n=\sqrt{\frac{G M_{\star}}{a^{3}}}\), indicating the rate at which the tidal forces act. 

To compare tidal heating with the flux \(F_{\mathrm{abs}}\) or to characterise the magnitude of tidal heating exhibited by a planet, we calculate the tidal flux from the tidal dissipation power \(\dot{E}_{\mathrm{tide}}\) using:  

\begin{equation} 
{F}_{\mathrm{tide}}=\frac{\dot{E}_{\mathrm{tide}}}{4 \pi R_{p}^{2}} 
\end{equation} 

Which represents the average energy flux produced by tides at the planet’s surface, expressed in \(\text{W m}^{-2}\), just like \(F_{\mathrm{abs}}\). 

\subsection{The Coexistence of Both Flows on the Same Exoplanet}

Stellar radiation is a heat source present in all planetary systems. However, for a planet to experience tidal heating, the condition \(e>0\) must be satisfied, with eccentricity being a necessary condition; although, once this property is satisfied, it acts primarily as a modulator of the flux \(F_{\mathrm{tide}}\) given its dependence \(F_{\mathrm{tide}} \propto e^{2}\). Whereas the semi-major axis \(a\) exhibits a stronger dependence, \(F_{\mathrm{tide}} \propto a^{-6}\), with \(a\) being dominant in the hierarchy based exclusively on the formula for \(\dot{E}_{\mathrm{tide}}\). Having explained both flows and mentioned the possible coexistence of these two within the same thermal environment, naturally, a comprehensive classification requires the formulation of a parameter that allows us to classify thermal regimes depending on the dominance exhibited by each exoplanet, as well as to compare physical mechanisms. 

\subsection{\(\Lambda\): Physical Parameter for the Classification of Thermal Regimes}

In this analysis, we use \(\Lambda\) as a dimensionless parameter that allows us to easily distinguish whether the thermal regime of an exoplanet is dominated by \(F_{\mathrm{abs}}\), \(F_{\mathrm{tide}}\), or whether both are comparable. We calculate the value of \(\Lambda\) using: 

\begin{equation} 
\Lambda=\frac{F_{\mathrm{abs}}}{F_{\mathrm{tide}}} 
\end{equation}  

And we organise the classification as follows: 

\begin{equation} 
\Lambda>1\to F_{\mathrm{abs}} \text{ dominance} 
\end{equation} 

\begin{equation} 
\Lambda<1\to F_{\mathrm{tide}} \text{ dominance} 
\end{equation}  

\begin{equation} 
\Lambda=  1\to F_{\mathrm{tide}} \sim F_{\mathrm{abs}} 
\end{equation} 

This allows us to reveal a fundamental hierarchy of the physical mechanisms influencing \(F_{\mathrm{abs}}\) and \(F_{\mathrm{tide}}\). Thus, we can order the physical regimes with respect to \(\Lambda\).

\section{Methodology}

In this study, we use a sample of approximately 2000 exoplanets whose observational parameters are taken from the “NASA Exoplanet Archive” \citep{Akeson2013}. 

We calculate the incident stellar flux \(F_{\mathrm{abs}}\) (see Section 2.2) using the formula:  

\begin{equation} 
F_{\mathrm{abs}}=\frac{L_{\star}}{4 \pi a^{2}}(1-A) 
\end{equation} 

We adopt a fixed value for the Bond Albedo \(A\): \(A=0.30\). 

We calculate the flux produced by tidal heating \(F_{\mathrm{tide}}\) (see section 2.3) using the expression: 

\begin{equation} 
\dot{E}_{\mathrm{tide}}=\frac{21}{2}\frac{k_{2}}{Q}\frac{GM_{\star}^{2}R_{p}^{5}}{a^{6}}ne^{2} 
\end{equation} 

Here, we assume the constants have the values: \(k_{2}=0.30\) and \(Q=100\) \citep{jackson2008, Henning2009, hansen2008}. The tidal dissipation power is converted to flux using: 

\begin{equation}  
{F}_{\mathrm{tide}}=\frac{\dot{E}_{\mathrm{tide}}}{4 \pi R_{p}^{2}}  
\end{equation}

We recall that the classification determined by \(\Lambda\) is calculated as the fraction of the incident stellar flux relative to the tidal flux. 

The analysis focuses on the classification of thermal regimes, the study of orders of magnitude, and the identification of relationships within the population. 

The calculations are performed in Python using standard libraries such as astropy, pandas, matplotlib and numpy.  

\section{Results}

\subsection{Transition between Tidal and Stellar Irradiation Regimes as a function of Orbital Period}

In this result we analyse the relative contribution of stellar radiation and tidal heating for different planetary systems, illustrating the transition between regimes dominated by radiation and those dominated by tides.

\begin{figure}[h!]
    \centering
    \includegraphics[width=1\linewidth]{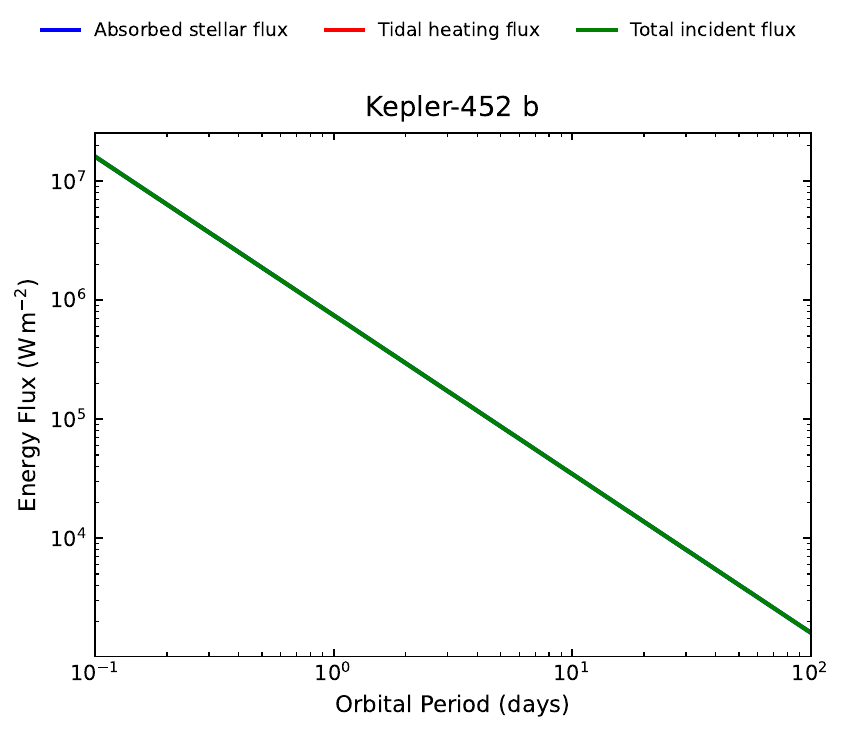}
    \caption{Energy fluxes as a function of orbital period for an irradiation-dominated case.}
    \label{fig:FirstResult1}
\end{figure}

\begin{figure}[h!]
    \centering
    \includegraphics[width=1\linewidth]{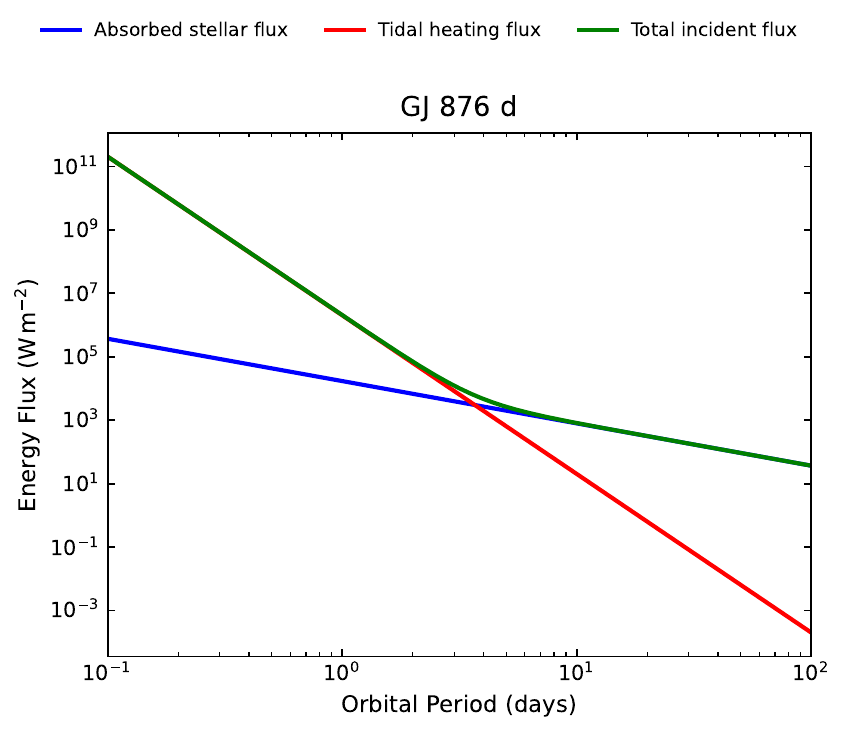}
    \caption{Energy fluxes for a tidally dominated case.}
    \label{fig:placeholder}
\end{figure}

These figures show a controlled comparison between two carefully selected planetary systems. Kepler-452 b is, to date, one of the exoplanets most similar to Earth; its orbit is circular, so, as it has no eccentricity, tidal heating is not a relevant heat source for this system, since there is no temporal variation in the tidal potential over its entire orbital period \(P\). In contrast, GJ 876 d is characterised by specific parameters that result in \(\Lambda<1\), and therefore tidal heating is the dominant source in the planet’s thermal environment. 

The figure corresponding to the exoplanet Kepler-452 b is an illustration where the incident stellar flux and the total incident flux are analogous (the green and blue lines overlap), which means that, as there is no tidal heating \(F_{\mathrm{tide}}\) (due to its parameter \(e=0\)), the thermal environment is characterised exclusively by the incident stellar flux \(F_{\mathrm{abs}}\).  

The remaining figure in this result, corresponding to the exoplanet GJ 876 d, is physically opposite due to its parameters favouring tidal heating, primarily \(e>0\) (a necessary but not sufficient condition) causes the periodic deformation of the planet, meaning that this is not the same at different points in its orbit, thus producing \(\Lambda<1\) and, consequently, tidal heating dominates the planet’s internal energy balance.

\subsection{Exoplanet Thermal Regimes Defined by the Flux Ratio \(\Lambda\)}

We begin our analysis with a comprehensive sample of exoplanets in order to characterise the overall distribution of thermal regimes. To this end, we use the dimensionless parameter \(\Lambda=F_{\mathrm{abs}} / F_{\mathrm{tide}}\), which quantifies the relative contribution of stellar radiation and tidal heating.

\begin{figure}[h!]
    \centering
    \includegraphics[width=1\linewidth]{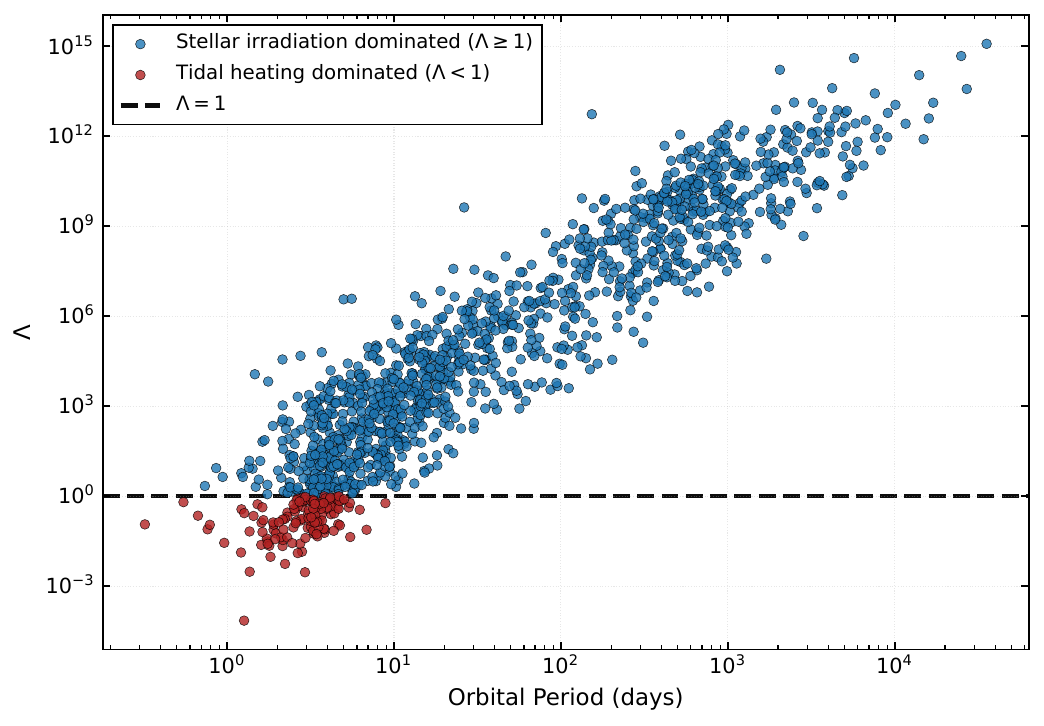}
    \caption{Thermal regimes of exoplanets defined by the flux ratio \(\Lambda\) as a function of orbital period. The blue points correspond to planets where radiation dominates (\(\Lambda>1\)), whilst the red points indicate that tidal heating dominates (\(\Lambda<1\)). The dashed line marks the transition between the two regimes (\(\Lambda=1\)).}
    \label{fig:SecondResult}
\end{figure}

This figure shows a diverse classification of 1869 planetary systems, whose parameters have been obtained from the catalogue used in this article. 

Each sphere, regardless of its colour, represents an exoplanet orbiting its host star.  

The figure is interpreted as follows: the Y-axis represents \(\Lambda\), which, as previously introduced, is a dimensionless parameter specific to this analysis that characterises the studied systems as falling within the ‘stellar irradiation domain’ or the ‘tidal heating domain’. The X-axis is the orbital period \(P\) of each planet, shown in days. These two classifications are separated by a dashed horizontal line representing \(\Lambda=1\), where both the blue and red spheres have a directly comparable stellar and tidal flux. The blue spheres dominate Fig.\ref{fig:SecondResult}, which leads us to conclude that the thermal environments of most of the planetary systems involved in this analysis are primarily dominated by the incident stellar flux \(F_{\mathrm{abs}}\). Apart from this dominance and below \(\Lambda=1\), there remains a significant number of exoplanets whose thermal environment is dominated primarily by tidal heating \(F_{\mathrm{tide}}\). 

This second result also illustrates a clear relationship between \(\Lambda\) and the orbital period \(P\), given that the higher the value of \(P\), the greater the dominance of the incident stellar flux \(F_{\mathrm{abs}}\) that characterises the exoplanets. Whereas we will scarcely find planetary systems dominated by tidal heating \(F_{\mathrm{tide}}\) for orbital periods \(P\) of 10 days or more.

\subsection{Dominant Parameters Governing Tidal Heating in Exoplanets}

To analyse the dependence of tidal heating, we plot \(F_{\mathrm{tide}}\) as a function of the main stellar, planetary and orbital parameters in the sample. This result shows \(F_{\mathrm{tide}}\) as a function of eccentricity, semi-major axis, planetary radius and stellar mass.

 \begin{center}
    \includegraphics[width=\linewidth]{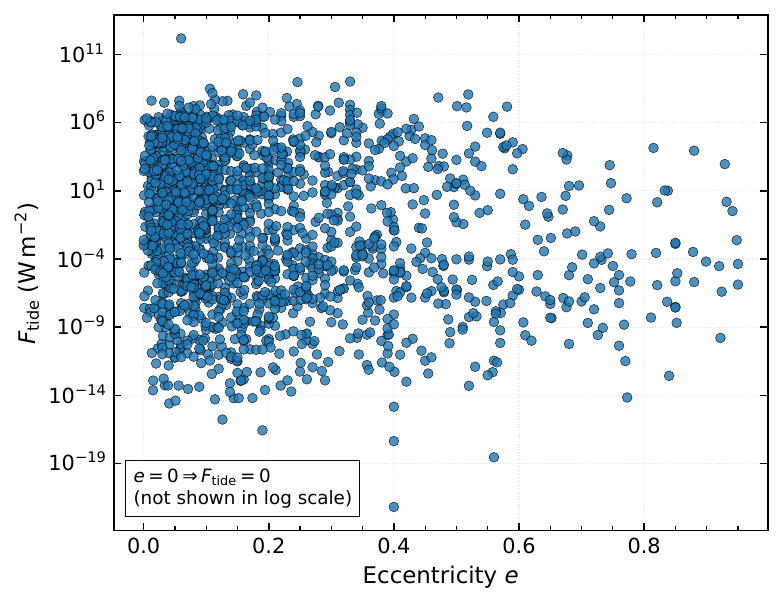}
    \captionof{figure}{Tidal heating flux \(F_{\mathrm{tide}}\) as a function of eccentricity \(e\) (cases with \(e=0\) are not shown in log scale).}
    \label{fig:ThirdResult1}
\end{center}

\FloatBarrier

 \begin{center}
    \includegraphics[width=\linewidth]{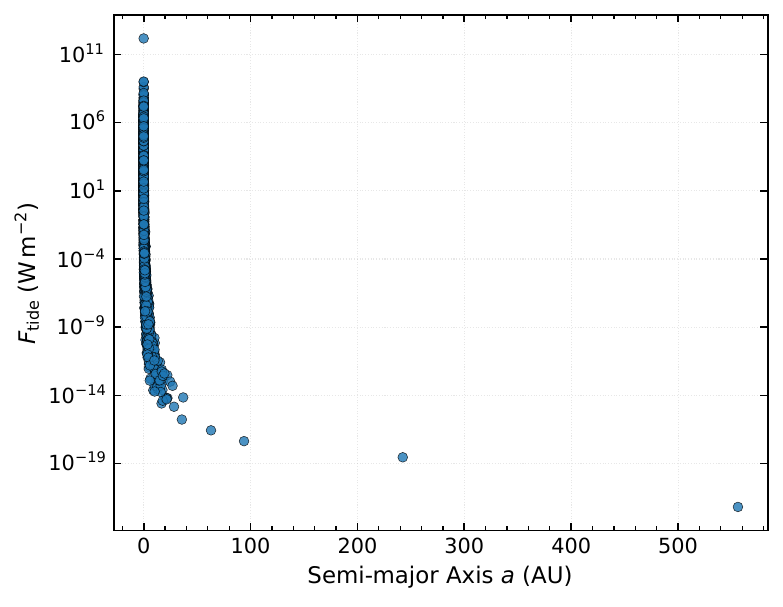}
    \captionof{figure}{Dependence of tidal heating flux \(F_{\mathrm{tide}}\) on semi-major axis \(a\).}
    \label{fig:ThirdResult2}
\end{center}

\FloatBarrier

 \begin{center}
    \includegraphics[width=\linewidth]{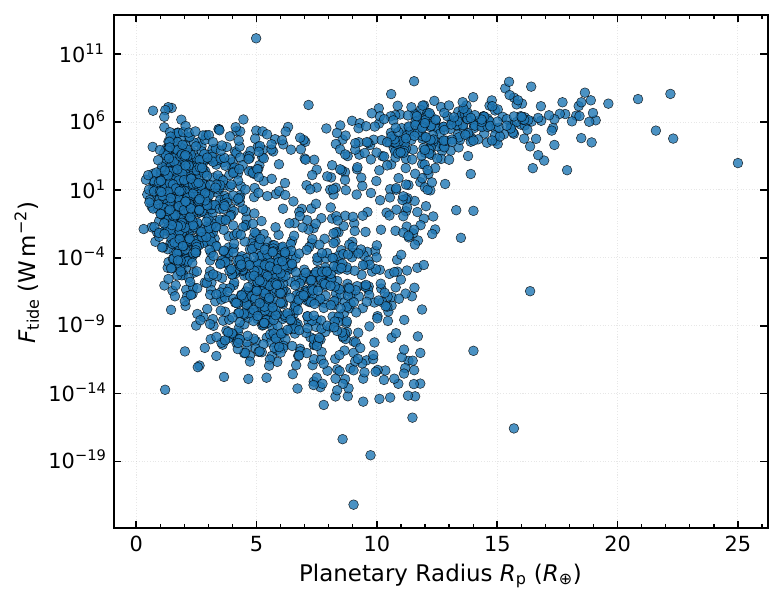}
    \captionof{figure}{Tidal heating flux \(F_{\mathrm{tide}}\) as a function of planetary radius \(R_{p}\).}
    \label{fig:ThirdResult3}
\end{center}

\FloatBarrier

 \begin{center}
    \includegraphics[width=\linewidth]{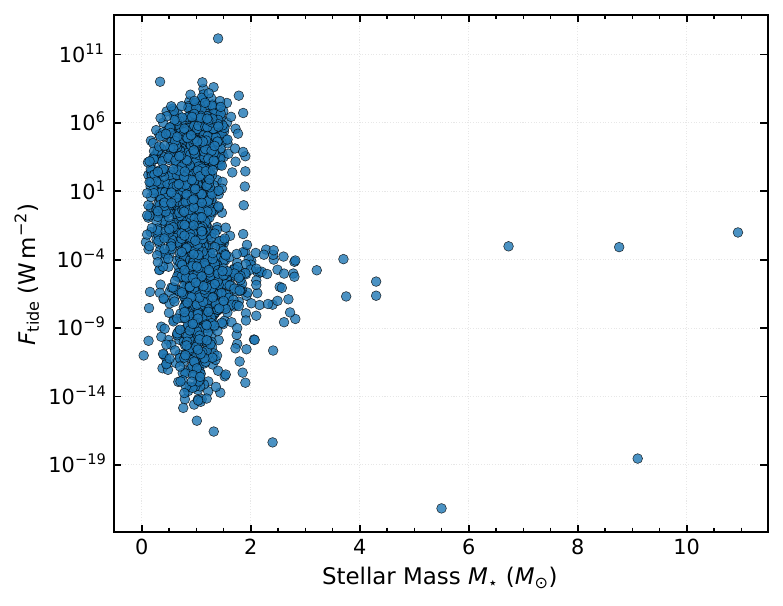}
    \captionof{figure}{Tidal heating flux \(F_{\mathrm{tide}}\) as a function of stellar mass \(M_{\star}\).}
    \label{fig:ThirdResult4}
\end{center}

Fig.~\ref{fig:ThirdResult1} compares the flow produced by tidal heating on the Y-axis against orbital eccentricity \(e\) on the X-axis. As each blue sphere represents an individual exoplanet whose parameters have been selected from the catalogue of preference for this article, their arrangement in the graph leads us to conclude that the necessary condition for the planet to experience tidal flow \(F_{\mathrm{tide}}\) is that the parameter shown on the X-axis must satisfy: \(e>0\), meaning that planetary systems with circular orbits do not experience tidal flow \(F_{\mathrm{tide}}\); this explains the absence of any tidal heating in the graph at \(e=0\), as highlighted in the caption for this figure. We conclude that \(e>0\) is a necessary condition for any tidal flux \(F_{\mathrm{tide}}\), although it is not sufficient on its own to determine its magnitude.

The second figure of the third result (Fig.~\ref{fig:ThirdResult2}) is a representation consistent with the strong dependence of \(F_{\mathrm{tide}}\propto a^{-6}\) that appears in the formula for the flow produced by tidal heating. The Y-axis retains the phenomenon studied in this result, \(F_{\mathrm{tide}}\), whilst the semi-major axis \(a\) now features on the X-axis. The figure illustrates the approximately 2000 exoplanets analysed in the third result, forming a marked concentration of planetary systems, mainly with small values of semi-major axis \(a\) between \(\sim\) \(0.01\) and \(40\) AU. Certainly, the dependence of \(a\) on \(F_{\mathrm{tide}}\) is strong; however, we cannot characterise the entire flux produced by tidal heating using only the semi-major axis \(a\). This confirms the point I mentioned earlier, namely that this rich variety of figures must be interpreted collectively to provide a rigorous characterisation of the importance of each parameter controlling \(F_{\mathrm{tide}}\).

In the third figure of the third result (Fig.~\ref{fig:ThirdResult3}), \(F_{\mathrm{tide}}\) once again takes centre stage on the Y-axis, but now it is the planetary radius \(R_{p}\) that occupies the X-axis and serves as the physical parameter to be studied in this figure. Although not to the same extent as the semi-major axis \(a\), the planetary radius \(R_{p}\) also constitutes an important factor in the tidal heating flux experienced by a planet; this relationship is represented as: \(F_{\mathrm{tide}}\propto R_{p}^{5}\). We observe a wide dispersion of planets in which the planetary radius \(R_{p}\) ranges from \(\sim 0.5\) to \(3\) \(R_{\oplus}\), resulting in high variability in \(F_{\mathrm{tide}}\). Furthermore, there is a notable tendency for higher fluxes produced by tidal heating \(F_{\mathrm{tide}}\) to occur in planetary systems where \(R_{p}\) is substantially higher. Again, this figure alone does not characterise the physical parameter under study in this result, \(F_{\mathrm{tide}}\), but rather we can classify the planetary radius \(R_{p}\) as a ‘physical amplifier’ that modulates the variability of the tidal heating flux \(F_{\mathrm{tide}}\) by an order of \(F_{\mathrm{tide}}\propto R_{p}^{5}\).

And finally, in the fourth figure (Fig.~\ref{fig:ThirdResult4}), the relationship is, once again, \(F_{\mathrm{tide}}\) on the Y-axis; and now, the stellar mass \(M_{\star}\) on the X-axis will be what orders our population of exoplanets with respect to the flux produced by tidal heating. The dependence shown here scales as: \(F_{\mathrm{tide}}\propto M_{\star}^{2}\), with \(M_{\star}\) acting as a modulation or amplification of \(F_{\mathrm{tide}}\). The figure illustrates virtually the entire population under study ranging between \(\sim 0.1\) and \(\sim 2\) \(M_{\odot}\); with a variation between \(\sim 10^{-14}\) and \(\sim 10^{8}\) \(\text{W m}^{-2}\). We conclude that the stellar mass \(M_{\star}\) is another parameter that must be considered in conjunction with others for a rigorous characterisation of the \(F_{\mathrm{tide}}\) phenomenon and its variability across different planetary systems.

\subsection{Two-dimensional plot of \(F_{\mathrm{tide}}\) versus \(F_{\mathrm{abs}}\)}

\begin{figure}[h!]
    \centering
    \includegraphics[width=1\linewidth]{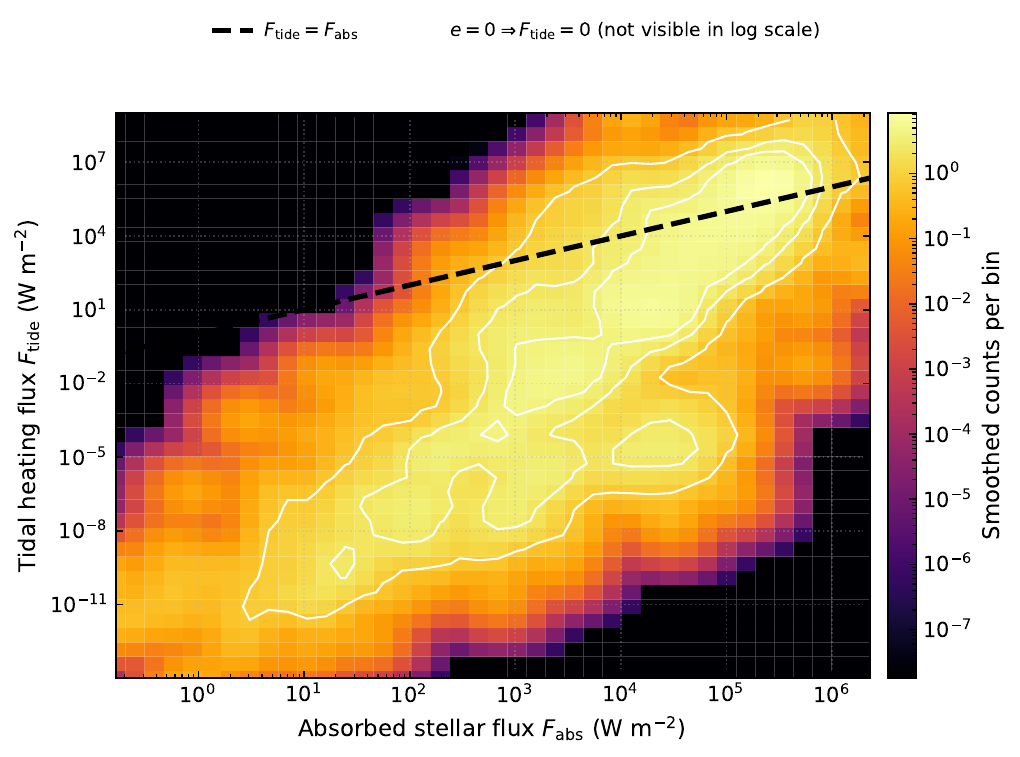}
    \caption{Two-dimensional plot of \(F_{\mathrm{tide}}\) versus \(F_{\mathrm{abs}}\). The dashed line marks the point where \(F_{\mathrm{tide}}=F_{\mathrm{abs}}\). The colours show the smoothed number of systems per interval in logarithmic space, and the contours indicate the regions of highest density.}
    \label{fig:FourthResult}
\end{figure}

Fig. \ref{fig:FourthResult} shows the two-dimensional distribution of the tidal heating flux \(F_{\mathrm{tide}}\) as a function of the absorbed stellar flux \(F_{\mathrm{abs}}\).

This new result depicts the analysed population on a density map. The exoplanets are distributed according to the two variables shown in the figure: \(F_{\mathrm{tide}}\) on the Y-axis, and \(F_{\mathrm{abs}}\) on the X-axis. Each bin or two-dimensional interval corresponds to a specific count of planets contained within that interval; the number of these corresponds to the colour of each pixel or cell, as defined in ‘smoothed count density per bin’. The dashed line crossing the figure precisely divides two physical regimes. The first regime (below the dashed line) is inhabited by exoplanets whose thermal environments are primarily dominated by the incident stellar flux \(F_{\mathrm{abs}}\). This is the most abundant population, with the various yellow zones located below the dashed line standing out as the most populated. The second regime (above the dashed line) is inhabited by a minority of planetary systems whose thermal environment is dominated mainly by the flux produced by tidal heating \(F_{\mathrm{tide}}\). Near the dashed line, both fluxes are of the same order, corresponding to \(\Lambda\simeq 1\), and therefore both fluxes are directly comparable. This density map does not show planetary systems with perfectly circular orbits, since, as we demonstrated in the previous result, tidal heating \(F_{\mathrm{tide}}\) only occurs in elliptical orbits where \(e>0\).

\subsection{Tidal Heating Landscape in the (\(a\), \(e\)) Parameter Space}

To isolate the orbital dependence of tidal heating, we exoplore the distribution of \(F_{\mathrm{tide}}\) in the (\(a\), \(e\)) parameter space.

\begin{figure}[h!]
    \centering
    \includegraphics[width=1\linewidth]{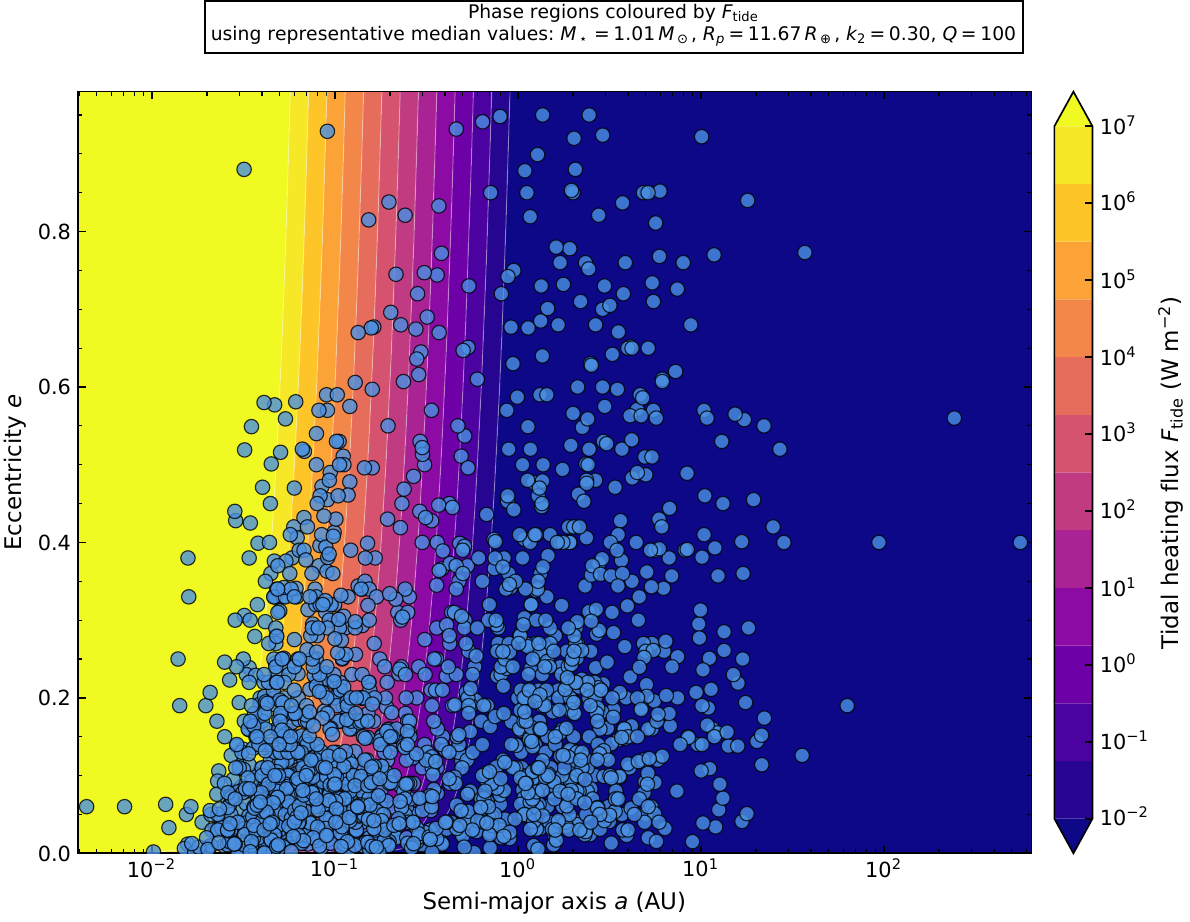}
    \caption{Tidal heating flux \(F_{\mathrm{tide}}\) across the (\(a\), \(e\)) parameter space. Colours indicate \(F_{\mathrm{tide}}\); points correspond to observed exoplanets, computed using representative median parameters.}
    \label{fig:FifthResult}
\end{figure}

This figure shows two of the parameters that govern tidal heating \(F_{\mathrm{tide}}\) and provides a more comprehensive population analysis. The coloured background represents different regimes of \(F_{\mathrm{tide}}\), whilst each blue point corresponds to an exoplanet located in the parameter space (\(a\) and \(e\)). A clear trend is observed: as the semi-major axis \(a\) decreases, the values of \(F_{\mathrm{tide}}\) increase, as indicated by the colour scale. The eccentricity \(e\) is necessary for tidal dissipation, since \(F_{\mathrm{tide}}\to 0\) when \(e=0\). For non-zero values, an increase in \(e\) leads to greater tidal flows, especially for small values of the semi-major axis \(a\). Overall, the figure shows that tidal heating is confined to a well-defined region of parameter space, where low values of \(a\) and non-zero eccentricities predominate.

\section{Discussion}

\subsection{General Physical Interpretation of the Results Obtained}

The results obtained in this analysis reveal the existence of a pattern of physical conditions that determine whether the thermal environment of a planetary system is likely to experience tidal heating, or whether such heating might even be so extreme as to give rise to an extreme atmospheric climate \citep{Barnes2009, Pierrehumbert2011}, thereby compromising the planet’s potential habitability. Among these stellar, planetary and orbital parameters, two of the most influential are: the semi-major axis \(a\), such that the shorter the star-planet distance, the greater the tidal fluxes \(F_{\mathrm{tide}}\) the system will experience; and the eccentricity \(e\), where the condition \(e>0\) must hold for tidal heating to occur on any scale, and the greater the \(e\), the greater the tidal fluxes, although it is not as critical a parameter as the semi-major axis \(a\). Therefore, these parameters should be prioritised when quantifying the fluxes \(F_{\mathrm{tide}}\); it is also worth noting the influence, albeit smaller, of the planetary radius \(R_{p}\) and the stellar mass \(M_{\star}\). Taken together, these results suggest that tidal heating is not a random process, but a predictable consequence of the orbital architecture and the physical properties of the system, which allows for the establishment of a hierarchical framework for interpreting the different thermal regimes in exoplanetary systems.  

\subsection{The influence of \(F_{\mathrm{tide}}\) on the Planetary Thermal Environment }

(Corresponding to Result I) In order to illustrate a significant change in the thermal environment of a planetary system caused primarily by the tidal flux \(F_{\mathrm{tide}}\), we have compared two exoplanets whose systems are vastly different and which clearly demonstrate this thermal consequence. The exoplanet Kepler-452 b (with characteristics similar to Earth) does not experience tidal heating due to its parameter \(e=0\), and, furthermore, its characteristic constant flux \(F_{\mathrm{tot}}\) reveals that the planet’s main thermal source is stellar \(F_{\mathrm{abs}}\), and shows no variations, as the body receives the same radiation at any point in its orbit. It is worth noting that its thermal environment and potential habitability depend primarily on the incident stellar flux \(F_{\mathrm{abs}}\). The exoplanet GJ 876 d differs radically in its characteristics compared to Kepler-452 b and, consequently, in its thermal environment and classification. This body does experience tidal heating \(F_{\mathrm{tide}}\) in addition to stellar radiation \(F_{\mathrm{abs}}\), resulting in higher and more variable values of \(F_{\mathrm{tot}}\). Given its parameter \(e>0\), it receives an incident stellar flux that is not constant throughout its orbit, and, compounded by its high eccentricity, a low value of the semi-major axis \(a\) causes the tides experienced by this planet to result in an extreme atmospheric climate, allowing us to classify it within our hierarchy as \(\Lambda<1\).  

\subsection{Population Classification based on our Variable \(\Lambda\)}

(Corresponding to Result II) As outlined above, this study introduces the parameter \(\Lambda\), which acts as a dimensionless indicator capable of quantifying which energy mechanism dominates the thermal environment of a planetary system. This has enabled us to carry out a comprehensive classification of approximately 2000 exoplanets and to determine the dominant energy mechanism in the thermal environment of each planet included in the analysis. As our results demonstrate, most planetary thermal environments are dominated primarily by the incident stellar flux \(F_{\mathrm{abs}}\), with stellar radiation being clearly dominant for orbital periods \(P\) of 10 days or more, suggesting that the orbital period acts as a controlling parameter of the thermal regime of planetary systems. Conversely, a minority (albeit significant) fraction of planetary systems are dominated by the flux \(F_{\mathrm{tide}}\), where stellar irradiation remains a relevant contribution and may give rise to extreme thermal conditions. Most exoplanets dominated by stellar radiation may also experience fluxes produced by tidal heating, although these are smaller in comparison to their dominant thermal source; exoplanets classified as \(\Lambda>1\) are less likely to have an extreme atmospheric climate and are therefore potentially more favourable for maintaining a stable thermal environment \citep{Kopparapu2014, spiegel2008}, a key characteristic in assessing a planet’s potential habitability. 

\subsection{Hierarchical Control of \(F_{\mathrm{tide}}\) in a Population of Planetary Systems}

(Corresponding to Result III) In this result, we present the fundamental hierarchy governing tidal heating \(F_{\mathrm{tide}}\) in a population comprising hundreds of planetary systems, which reveals the observed trends. The eccentricity \(e\) is a key parameter, as the condition \(e>0\) excludes all systems with circular orbits from experiencing any flux produced by tidal heating, although it does not act as a dominant parameter in the population; this parameter constitutes an essential condition for a thermal environment to experience the tidal flux \(F_{\mathrm{tide}}\). The semi-major axis \(a\) is the dominant parameter for analysing the strength of tidal flows; our graphs reveal that the smaller the value of \(a\), the greater the tidal heating-induced flows a planet may experience, provided other parameters of importance for determining the influence of \(F_{\mathrm{tide}}\) on the thermal environment are present. This parameter in the hierarchy is certainly the one with the greatest influence on the tidal flux, given its physical dependence \(F_{\mathrm{tide}} \propto a^{-6}\). The other two parameters comprising this hierarchy are less influential in characterising the order of magnitude of \(F_{\mathrm{tide}}\), although they are not negligible. One of these is the planetary radius \(R_{p}\), whose trend in our results shows that the higher the values of \(R_{p}\), the greater the \(F_{\mathrm{tide}}\) flows characterising the planet’s thermal environment. It is worth mentioning the physical relationship \(F_{\mathrm{tide}}\propto R_{p}^{5}\), which justifies the importance of considering this parameter within the hierarchy in order to conduct a consistent analysis of the degree of dominance of tidal heating in the thermal regimes of planetary systems. And the final parameter that we must not overlook in the hierarchy is the stellar mass \(M_{\star}\); in our figures, an indirect correlation can be observed in comparison with \(a\), leading to the conclusion that its effect on the variation of \(F_{\mathrm{tide}}\) is secondary compared to the previous parameters. Taken together, these parameters—some dominant and others modulating—form a fundamental hierarchy that allows us to classify the planets under analysis according to which energy mechanism dominates their thermal environment, as confirmed by plotting a large population of planetary systems on a density map in the subsequent result. 

\subsection{Thermal Phase Space in Exoplanets: The Role of \(F_{\mathrm{tide}}\) and \(F_{\mathrm{abs}}\) Flows}

(Corresponding to Result IV) We analyse jointly the influence of the incident stellar flux and the tidal flux on the thermal environment of a planetary system using a density map applied to the population of planets under study. A trend is observed between the two fluxes, such that as \(F_{\mathrm{abs}}\) increases, \(F_{\mathrm{tide}}\) is associated with higher values of \(F_{\mathrm{tide}}\). We identify a physical boundary represented by \(F_{\mathrm{tide}}=F_{\mathrm{abs}}\) that separates the two regimes. Systems dominated primarily by \(F_{\mathrm{abs}}\) correspond to \(\Lambda>1\), whilst those dominated by \(F_{\mathrm{tide}}\) correspond to \(\Lambda<1\). For planetary systems closest to this conceptual boundary where \(\Lambda\simeq 1\), both fluxes can be considered comparable. This map does not include a significant fraction of the population, as these systems have perfectly circular orbits (\(e=0\)) and lack a tidal flow to represent them, which highlights that tidal heating \(F_{\mathrm{tide}}\) is not universal, but conditional. This dispersion can be explained by the dependence of \(F_{\mathrm{tide}}\) on additional parameters that modulate \(F_{\mathrm{tide}}\), such as the eccentricity \(e\), the internal structure of the planet (\(k_{2}\), \(Q\)), the stellar mass \(M_{\star}\) or the planetary radius \(R_{p}\). In summary, this map represents the population of planetary systems where \(e>0\) as a function of the energy fluxes they experience. Although most thermal environments are characterised by \(\Lambda>1\), there is a minority but significant population where \(\Lambda<1\) and which satisfy specific conditions capable of giving rise to extreme thermal environments. 

\subsection{Mapping of Regimes Established by Tidal Flow in the Orbital Phase Space}

(Corresponding to Result V) In this final result, we present a map ordered according to the tidal flow \(F_{\mathrm{tide}}\) with respect to the two parameters that most influence the flow: the eccentricity \(e\) and the semi-major axis \(a\). As we stated in the result setting out the proposed fundamental hierarchy, the eccentricity \(e\) is primarily a condition; once this is met, it becomes a secondary modulator with respect to the semi-major axis \(a\), the dominant parameter of the hierarchy. The figure representing the fifth result now reveals, geometrically and with greater precision, the trend of the vertical distribution followed by planetary systems when modulated by \(e\), and the virtually complete characterisation that \(a\) imposes on the tidal flow horizontally, as shown in the diagram. What this characterisation provides is precisely where the real systems ‘fall’ within the theoretical space, revealing the environment in which all relationships coexist. 

\section{Conclusions}

In this paper, we present a reproducible physical framework with the aim of classifying the order of magnitude of the tidal flux and studying, at a population level, its influence on the thermal regimes of planetary systems. 

Our framework reveals which energy mechanism is dominant in systems with different stellar, planetary and orbital parameters, using the characterisation proposed by \(\Lambda\). Within this framework, we identify a fundamental hierarchy of physical mechanisms that must be taken into account when calculating the value of \(F_{\mathrm{tide}}\). Furthermore, this hierarchy has been studied in a population of approximately 2000 planetary systems with different characteristics, revealing a non-trivial trend.  

The results of our paper are summarised as follows. All exoplanets experience stellar irradiation, and this energy mechanism dominates the thermal regimes of the majority of the studied population. A non-negligible minority experiences the tidal flux \(F_{\mathrm{tide}}\), which is dominant in the thermal environment if \(\Lambda<1\) or competes with \(F_{\mathrm{abs}}\) if \(\Lambda=1\). Planetary systems characterised by \(\Lambda>1\), where the incident stellar flux dominates over tidal heating, indicate that these systems do not present favourable conditions for tides to dominate the radiative balance. One of the most influential parameters on \(F_{\mathrm{tide}}\) is the eccentricity \(e\). The condition \(e>0\) is necessary for a planet to experience tidal heating. When this condition is met, the eccentricity becomes primarily a second-order modulating factor compared to the semi-major axis \(a\). The star-planet distance or semi-major axis is the dominant parameter in the hierarchy, as the physical dependence \(F_{\mathrm{tide}} \propto a^{-6}\) highlights the importance of \(a\) on the tidal flux.  

In physical terms, we can highlight that the thermal environment of an exoplanet is not purely radiative, but may be hybrid when both fluxes coexist. The framework presented establishes a fundamental hierarchy of four essential parameters for a comprehensive characterisation of the variability of the tidal flux \(F_{\mathrm{tide}}\). We conclude that studying tidal heating is relevant given the existence of hybrid planetary systems whose thermal environment experiences both fluxes, highlighting distinct thermal regimes. 

All our results, including the proposed framework, have been presented in a population analysis, demonstrating a non-trivial trend consistent with our explanations. Furthermore, in the figures we have identified planetary systems with extreme thermal regimes consistent with the physical conditions of their orbital, planetary and stellar parameters. We also emphasise the importance of the \(F_{\mathrm{tide}}\) flux when analysing a planet’s thermal environment, as this flux can produce unexpected variations in thermal regimes and thus affect potential habitability.

This analysis has certain limitations, such as the simplification of the physical model to a simple energy balance, as we reduce the characterisation of the thermal environment to the fluxes \(F_{\mathrm{abs}}\) and \(F_{\mathrm{tide}}\). We do not explicitly consider atmospheric effects, such as the greenhouse effect or a variable Bond albedo. All parameters used in the analysis have been selected from observations in our preferred catalogue, which contain uncertainties; therefore, the accuracy of the results depends on the quality of the data. Tides depend considerably on the planet’s internal structure; in this analysis, the values \(k_{2}=0.30\) and \(Q=100\) are set as defaults when calculating any flux \(F_{\mathrm{tide}}\) or \(F_{\mathrm{abs}}\). The study is based on a population analysis, which does not involve specific modelling of individual systems or detailed simulations; we therefore focus on the detected and demonstrable trends. In several figures, systems with circular orbits where \(e=0\) are excluded, as they do not exhibit tidal fluxes. Our classification parameter \(\Lambda\) does not capture temporal effects, nor does it include full orbital variability; rather, it is reduced to a general ratio. The exoplanets comprising the studied population correspond mainly to nearby systems, which reduces uncertainties but is not fully representative of the real universe. Energy sources such as internal radioactivity or gravitational contraction are reduced to the fluxes under analysis. 

Future research may include the flux produced by tidal heating in more complex atmospheric models or apply \(F_{\mathrm{tide}}\) to specific subpopulations. It may even link to future observations using data from more distant systems. 

In conclusion, this study provides a unified perspective on planetary thermal environments and establishes a conceptual framework for interpreting the diversity of observed systems.  

\newpage

\bibliographystyle{aa}
\bibliography{references}

\end{document}